\documentclass[twocolumn,twoside,slac]{revtex4}
\usepackage{psfig}
\usepackage{graphicx}
\usepackage{fancyhdr}
\renewcommand{\headrulewidth}{0pt}
\renewcommand{\footrulewidth}{0pt}

\def\Journal#1#2#3#4{{#1} {\bf #2}, #3 (#4)}

\def\NIMA{{\em Nucl. Instrum. Methods} A}

\newcommand{\beq}{\begin{equation}}
\newcommand{\eeq}{\end{equation}}
\newcommand{\z}{\mbox{$z$}}
\newcommand{\zxt}{\mbox{{\rm \z-by-timing}}}
\newcommand{\rphi}{\mbox{$r$-$\phi$ }}

\begin{document}

\title{The Architecture of the ZEUS Micro Vertex Detector 
DAQ and Second Level Global Track Trigger}

\author{A. Polini}
\affiliation{DESY, Notkestr. 85, 22607 Hamburg, Germany. E-mail: alessandro.polini@desy.de}

\begin{abstract}

The architecture of the ZEUS Micro Vertex Detector data acquisition system 
and the implementation 
of its second level trigger, the ZEUS Global Track Trigger are described. 
Data from the vertex detectors HELIX read-out chips, corresponding to 200k
channels, are digitized by 3 crates of ADCs which perform noise and pedestal 
subtraction, and data suppression and compaction. PowerPC VME board computers 
push cluster data for second level trigger processing and strip data for event 
building via Fast and Gigabit Ethernet network connections. Additional tracking 
information from the central tracking chamber and the forward 
straw tube tracker 
are interfaced into the 12 dual CPU PC farm of the Global Track Trigger where 
track and vertex finding is performed. The system is data driven at the ZEUS 
first level trigger rate ($\sim$ 500Hz) and must generate a trigger result 
after a mean time of 10ms.  

\end{abstract}

\maketitle

\thispagestyle{fancy}

\section{Introduction}\label{sec:intro}

The ZEUS detector at DESY is designed to study
high energy interactions produced at the HERA $e^{\pm} p$ collider. 
In the period 2000-2001, the 920 GeV proton - 27 GeV electron collider 
underwent a substantial upgrade aimed at increasing by a factor 5 
the peak luminosity, corresponding to 200 pb$^{-1}$ integrated luminosity 
per year. 
During the upgrade shutdown, ZEUS has been equipped with a 
silicon vertex detector which, besides a general improvement and extension 
of the track reconstruction, will enhance the identification of short lived 
particles. 

This paper describes in sections 1-8 the hardware 
architecture of the Data Acquisition 
system and the embedded Global Tracking Trigger (GTT). Section 9-11 outline 
the software solution used. Section 12 describes the trigger algorithm. 
Performance and Outlook are reviewed in sections 13 and 14. 

\section{Detector Layout}\label{sec:lay}

\begin{figure*}
\hbox{
\hspace*{-0.4cm}
\raisebox{-0.12cm }
{\psfig{figure=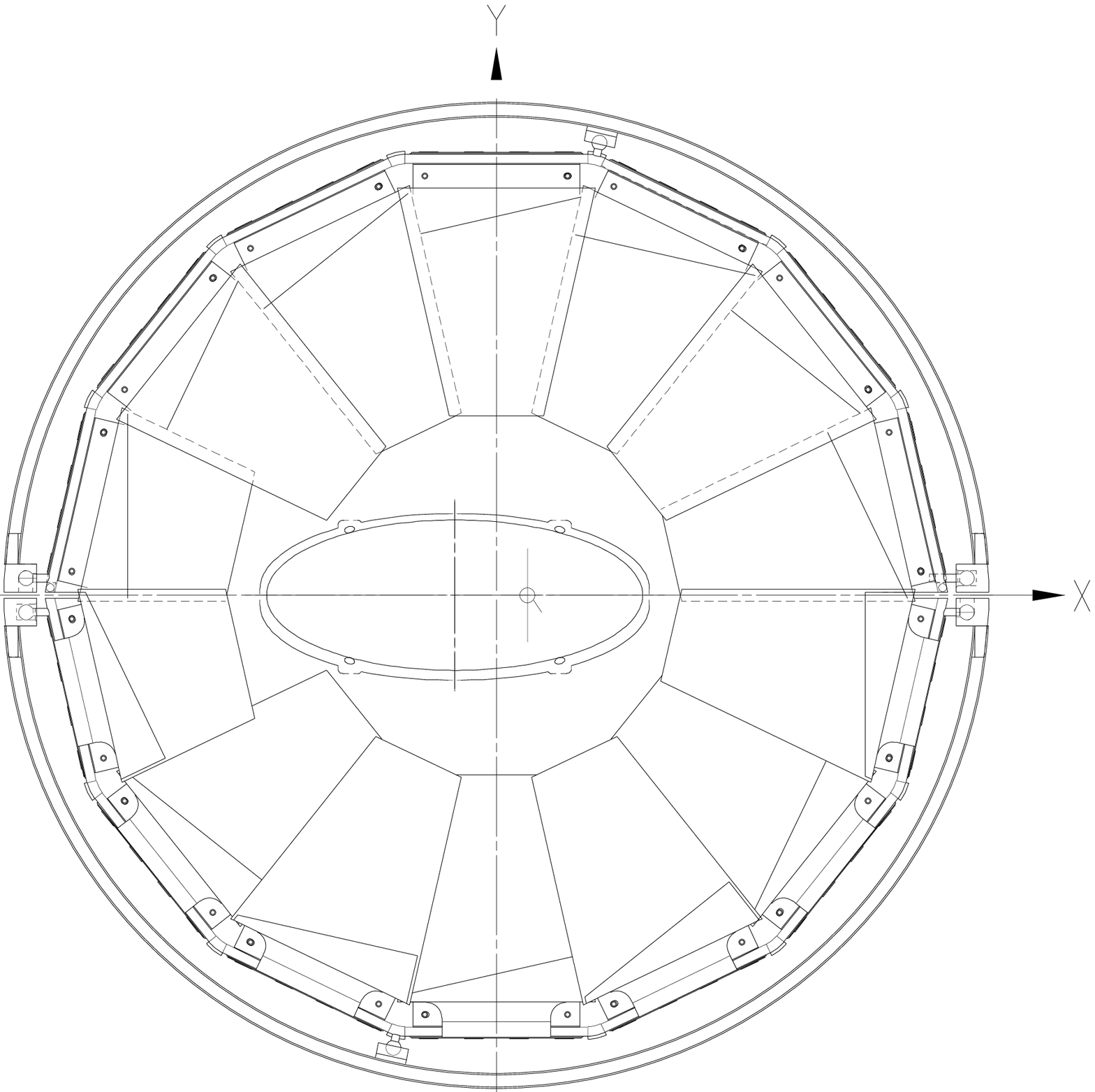,height=3.cm,clip=}} 
\hspace*{-0.9cm}
\raisebox{2.8cm}[0.cm][0.cm]{$ (a)$}
\hspace*{0.1cm}
\raisebox{-0.2cm }
{\psfig{figure=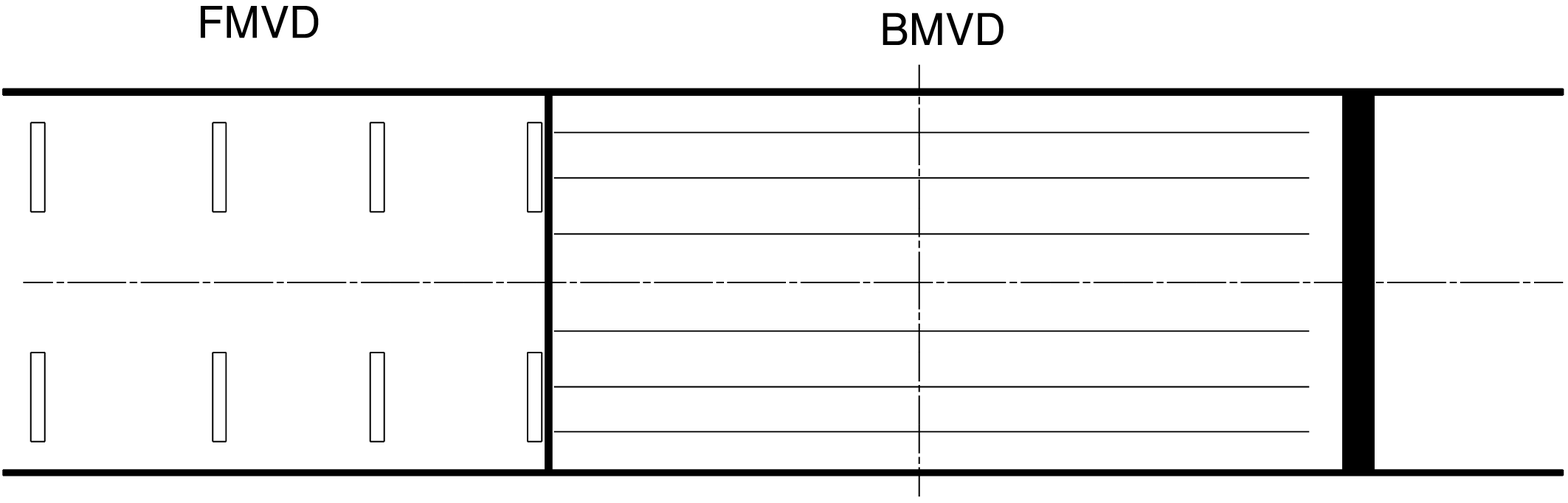,height=3.cm,clip=}} 
\hspace*{-0.5cm}
\hspace*{-0.5cm}
\raisebox{2.8cm}[0.cm][0.cm]{$ (b)$}
\hspace*{0.1cm}
\raisebox{-.23cm}
{\psfig{figure=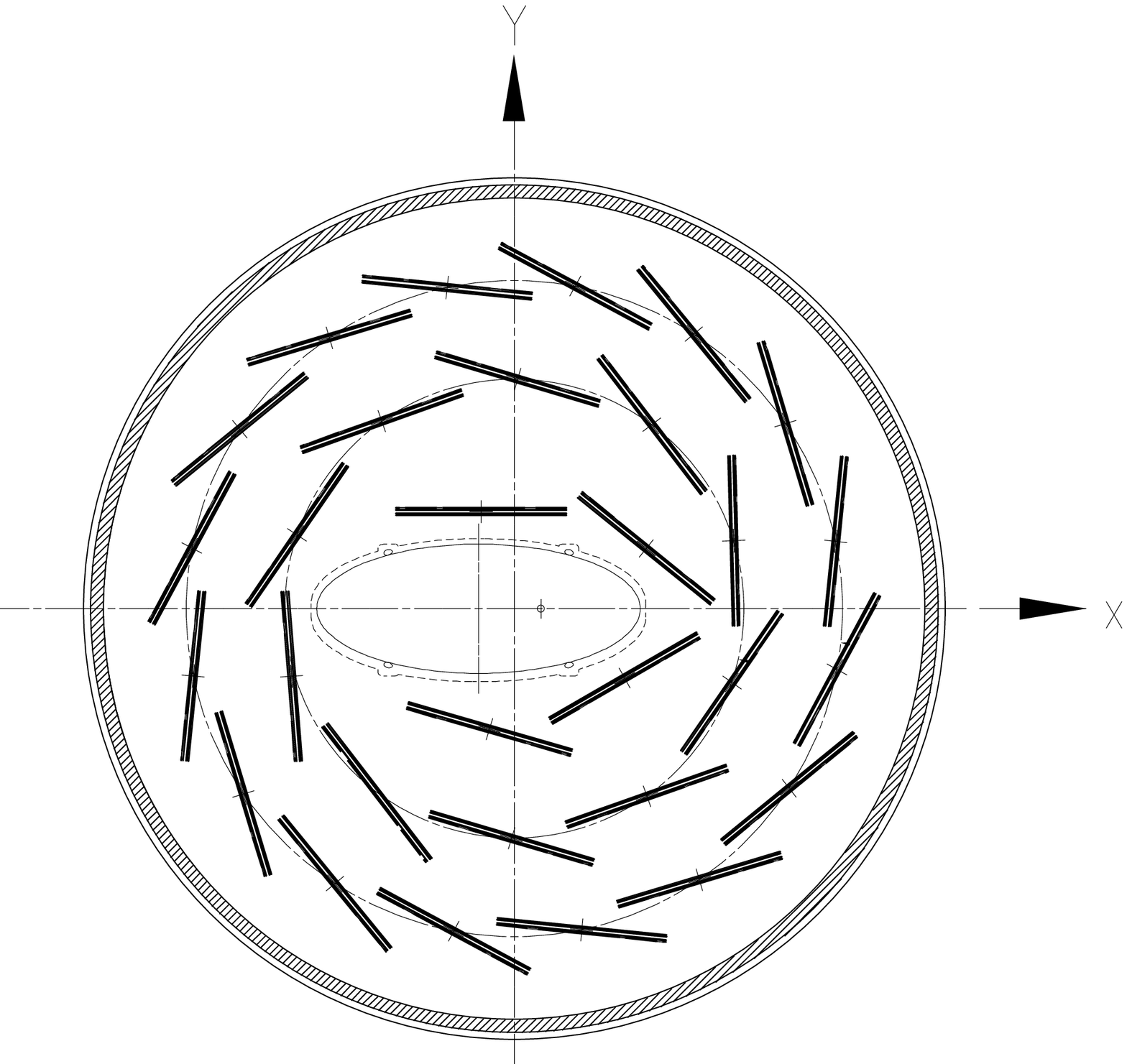,height=3.5cm,clip=}} 
\hspace*{-1.0cm}
\raisebox{2.8cm}[0.cm][0.cm]{$ (c)$}
\hspace*{-13.2cm}
\raisebox{2.8cm}[0.cm][0.cm]{Forward MVD}
\hspace*{2.5cm}
\raisebox{2.8cm}[0.cm][0.cm]{ Barrel MVD}
}
\caption{\label{fig.layout} Transversal view with respect to the beam line 
of the forward (a) and the barrel sections (c) and longitudinal 
view (b) of the complete detector.
\vspace{-0.1cm}
}
\end{figure*}

The Micro Vertex Detector (MVD) consists of a barrel section with 
three double layers of silicon sensors surrounding the beampipe and four 
wheels in the forward, outgoing proton, direction.
Longitudinal and transversal views of the detector with respect to the 
beam line are shown in Fig.~\ref{fig.layout}.  

The sensors are single sided and made of high-resistivity 
($3-6$\,k$\Omega$\,cm) 320 $\mu$m thick n-type silicon into which 
$p^+$ strips, 12 $\mu$m wide and with a 20 $\mu$m pitch, are implanted. 
The signal is read out via AC coupling of 14 $\mu$m strips placed at a pitch 
of 120 $\mu$m. The rear side consists of a thick $n^+$ diffusion.
Test beam results have shown that, using capacitive charge sharing, a 
resolution up to 8 $\mu$m can be obtained for tracks perpendicular to the 
sensor. 

In the barrel region 
two consecutive sensors of square shape ($60 \times 60$ mm), with 
orthogonal strips, are glued and electrically connected together
via a copper 
trace etched on 50 $\mu$m thick Upilex foil. 
The connection of the sensor 
assembly to the read-out hybrid also uses Upilex foil. 
This structure with a mirror one having perpendicular strip orientation 
forms a barrel {\it module} with 2048 read-out strips or 1024 channels.
Five modules are mounted on a carbon fiber ladder that provides 
the required stiffness and support for the cooling pipes, cabling and 
slow control sensors.

The forward section consists of four wheels, each made of two parallel
layers of 14 silicon sensors of same type as the barrel section but with 
a trapezoidal shape and 480 read-out channels. 
Two sensors mounted  behind each other form a {\it forward segment} and 
provide a two coordinate measurement via strips tilted by $180^\circ/14$ 
in opposite directions. A more detailed description of 
the detector layout and of the silicon sensors can be found 
in~\cite{layout}.

\section{The MVD Front-end and Read-out Electronics}\label{sec:readout}

The read-out of the 207,360 channels is performed by the {\tt HELIX\,128-3.0} 
front-end chip~\cite{helix}, a 0.8 $\mu$m CMOS chip 
specifically designed for the HERA environment. 
Each of the 128 channels is equipped with a preamplifier, 
a shaper and an analog 136 step analog pipeline. A pipeline read-out amplifier, 
a 40 MHz multiplexer and a 40 MHz current buffer form the backend 
stage of the design. 
The noise performance of the chip depends on the input 
capacitance (C) and is $400 + 40 \cdot$ C [pF] equivalent noise charge (ENC).
The bias settings and various other parameters of the analog read-out 
can be finely adjusted to optimize the system to the detector input 
characteristics and correct for radiation damage effects. 
Irradiation tests  of the {\tt HELIX} read-out
have been performed and indicate
that a total dose of 300 kRads can be received before degrading the
performance. The anticipated dose of 10-20 kRads per year
at HERA will allow more than the 5 years of operation foreseen. 

During operation the power dissipation of the {\tt HELIX} 
is 2mW per channel while the power dissipation in the silicon 
sensors is negligible. 

The bias setting and register programming as well as the 
clock, trigger and test pulse synchronization, are performed 
via a serial interface driven by custom designed 6U VME driver boards. 
Eight chips belonging to the same barrel module or the same forward segment 
are connected together in a
programmable {\it failsafe token ring}\footnote{
By providing 2 inputs and 2 output connection lines from each chip 
to its neighbours, the failsafe token ring, 
allows any subchain with not more than one consecutive faulty chips 
to be read out. This reduces the impact of faulty chips on the number 
of channels read out per chain.} 
and read out via a single digitization channel. 

The analog serialized output data are sent through passive copper 
links to dedicated 10 bit ADC VME embedded modules~\cite{adcm}. 
At the input to the
ADC module the signal range is 0-2V corresponding to 0-10 minimum 
ionizing particle. The ADC
system performs common mode noise and pedestal subtraction and writes
data into {\it strip} or {\it cluster} cyclic event data buffers. 
The strip buffer contains either
raw or strip data where an energy threshold cut is applied. 
The cluster buffer contains information from groups of contiguous strips 
 (center value, total energy, first and last strips,
etc.); a cut can be applied on the total energy and dead or hot
channels can be masked. Cluster data is used for triggering purposes
and  strip/raw data for event builder read-out.
The ADC system has been implemented as a 9U single width VME board each 
with 8 analogue inputs.  The total of 206 tokens are distributed in three
VME crates: upper-barrel, lower-barrel and wheels.

The {\tt HELIX} front-end and ADCs are controlled by the 
{\it clock-and-control} system. This consists of a single {\it master} with
three ADC crate {\it slaves} and the {\tt HELIX}-{\it driver} system. 
The clock-and-control system provides the interfaces between the ZEUS
{\it Global First Level Trigger} (GFLT), the run control system, the ADC crates and
the front-end chips. The {\tt HELIX}-driver boards 
are used by the run control to configure the system and by the master 
to propagate
trigger accepts to the {\tt HELIX} which outputs the analogue signal
with its own scheduling after receiving the trigger. The system is 
free running, the ADC modules inhibiting the trigger (busy) when the 
data buffer full condition occurs. 
The ADC system performs several checks to ensure correct event
processing. The length 
of the input data is checked
for its correspondence to the configured number of data strips
expected. The cell number is decoded from trailer data to 
identify pipeline jumps. Data arrival times are measured with respect to the
trigger timing. Error conditions are written into the output buffers
and fatal errors can assert a master error blocking the trigger.   

The ADC read-out is performed via VME PowerPC boards running LynxOS 3.01
and a dedicated software library~\cite{uvmelib}. 
The decision to use Motorola PowerPC read-out CPUs was made in 1999
after feasibility studies on the VME and network data transfer bandwidth 
and latency measurements. The use of LynxOS, at that time, was essential 
as it provided a UNIX environment with a realtime kernel. Priority scheduling
of the interrupt handling, VME data read-out via independent DMA and network 
transfer task pipeline were required to reach the necessary performance. 

\section{The ZEUS Data Acquisition System}

The ZEUS data acquisition system~\cite{zeus93} is based on a three level 
trigger. 
Because of the HERA bunch crossing rate of 10.4 MHz, i.e. 96 ns between 
consecutive interactions, the experiment is required to use a pipelined 
read-out design. The GFLT, based on a reduced 
set of information from the detector components, is issued after 46 bunch 
crossings and reduces the trigger rate to $\le 500$ Hz. 
Detector data, stored in deadtime-free analog or digital pipelines, 
is subsequently digitized, buffered and used by the {\it Global Second Level 
Trigger} (GSLT). The GSLT lowers the trigger rate to $\leq$70 Hz
with a typical latency of 10-15 ms. 
For accepted events the complete detector information is read out, 
merged with the other detector components data by the {\it Event Builder}
(EVB) and sent to the {\it Third Level Trigger} (TLT) computer farm where
event reconstruction and final online selection are performed. 
In normal data taking conditions the system runs with a deadtime 
$<2$\%.

\begin{figure*}[htb]
{\centerline{\psfig{figure=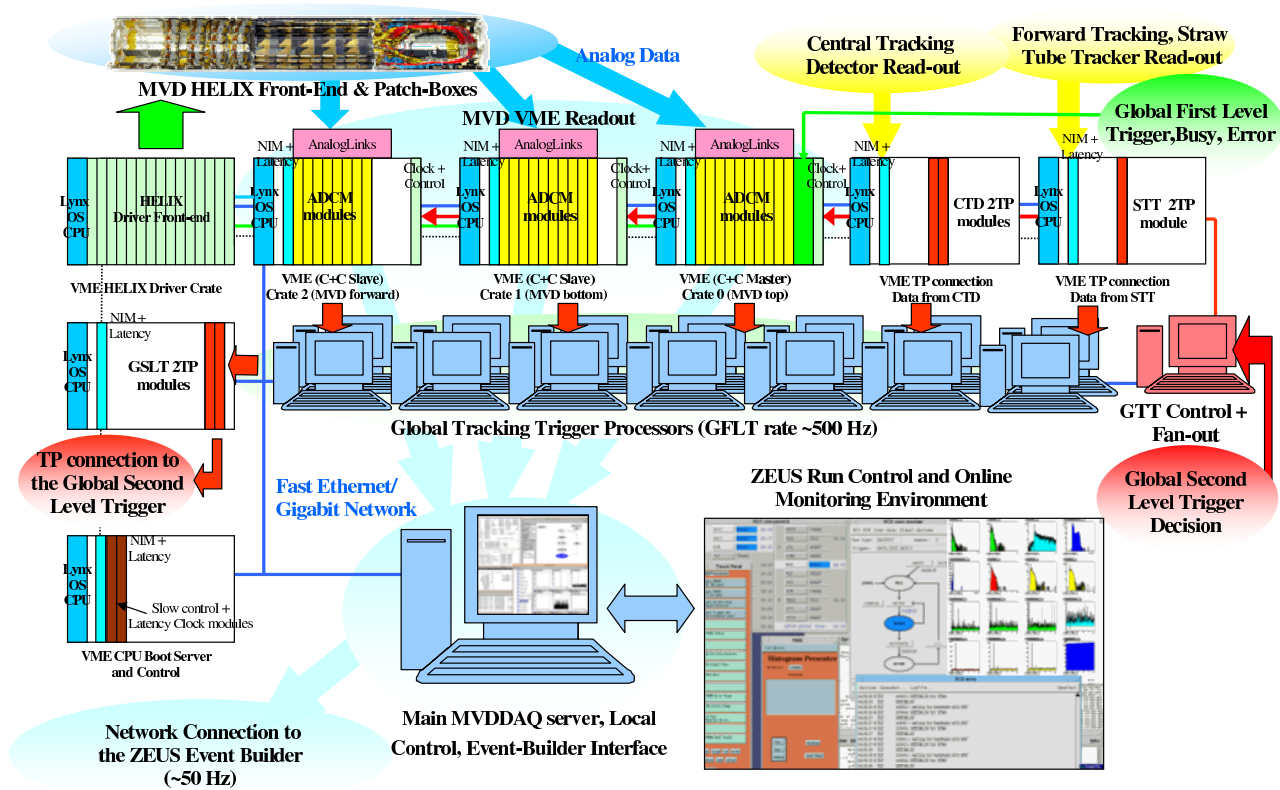,width=17cm,clip=}}}
\caption{Hardware implementation of thew Micro Vertex Detector
Data Acquisition system and the embedded GTT. }
\label{fig.mvddaq}
\end{figure*}

\section{The Global Tracking Trigger}

The architecture of the MVD DAQ has been strongly influenced by the
development of its contribution to the ZEUS trigger. 
As the {\tt HELIX} front-end and ADC read-out is too slow to participate 
to the First Level Trigger a contribution at the second level was targeted. 

Simulation studies on data multiplicity and background have shown 
that the MVD data alone, providing up to 3 planes of information per track 
would not be sufficient for unambiguous tracking and efficient rate 
reduction. Combining MVD information 
with the surrounding tracking detectors allowed a far better rate 
reduction and tracking efficiency. 
These considerations have lead to the design 
of the new Global Track Trigger, a distributed computing environment 
processing data from MVD, the existing Central Tracking Detector (CTD)
and the newly installed forward Straw Tube Tracker (STT).  

The CTD~\cite{ctd} is a cylindrical drift chamber surrounding 
the MVD and covering the polar angle range 
$15^\circ < \theta < 164^\circ$. It consists of 72 radial layers grouped  
into 9 superlayers: the odd superlayers have axial wires parallel to the beam 
axis while the even superlayers have small angle $(\pm 5 ^\circ)$ stereo 
wires allowing determination of the $z$ position of hits. 
The three inner axial layers are also equipped 
with $z$-by-timing electronics which determine the $z$-position of a hit 
from the difference in arrival times of a pulse at both ends of the chamber.

The STT~\cite{stt} consists of 4 superlayers of straw drift tubes of 7.75 mm 
inner diameter covering the polar angle $6^\circ < \theta < 23 ^\circ$. 
Each superlayer contains 2 planes composed of 6 trapezoidal 
shape sectors covering the full azimuthal angle. 
A sector consists of 3 vertical layers of straw tubes oriented in the 
azimuthal direction and providing an accurate measurement of 
the radial coordinate of the tracks. 
The STT superlayers are hosted in groups of two between plane 1 and 2 
and plane 3 and 4 of the existing ZEUS Transition Radiation Detector. 

In the design of the DAQ and GTT system, the preferred choice has been to 
use, whenever possible, commercial off the shelf equipment easily 
upgradeable and maintainable. After investigation of the performance achievable 
in terms of data throughput, process latency and performance, a solution based 
on a farm of standard PCs connected via a Fast/Gigabit Ethernet network 
has been chosen. 

The final hardware implementation 
of the MVD DAQ and the GTT systems is shown in Fig.\ref{fig.mvddaq}. 
In table~\ref{tab.hw} the hardware characteristics are listed.

On GFLT accept data is transferred from the silicon detector {\tt HELIX} 
read-out chip
pipelines to $\sim$30 ADC boards where data is digitized 
and stored in strip and cluster FIFO buffers. 
On completion of digitization a VME interrupt is generated and 
the cluster data is read out, via VMEbus, by a read-out CPU (Motorola MVME 2400
PowerPC/LynxOS) and sent to a GTT reconstruction environment. 
The GTT result is then forwarded to the GSLT interface 
hosted in another VME system. If the event is accepted at 
GSLT the complete read-out of the strip/raw data from the 3 MVD crates 
together with the complete GTT output is sent to the event building
process.

\section{The CTD and STT Interfaces}

The CTD and STT DAQ systems, as many other component in the 
ZEUS experiment, are based on transputers
\footnote {INMOS Transputers, were an advanced technological
development in the early 90's when the ZEUS experiment was designed.
Provided with a 32 bit processing unit, on board memory, four 20~MHz 
serial links for processor interconnection and a high level parallel 
programming language (OCCAM), transputers were ideal for highly 
distributed parallel processing and data transfer.}. 
In order to connect these
component to the GTT, the existing transputer networks had 
to be extended and an interface from the transputer protocol to the 
plain Ethernet network was required. 
At the time of design no commercial solution with the required 
flexibility and bandwidth was available; a VMEbus system based on the 
same read-out CPUs used in the MVD was commissioned~\cite{ctdint}. 
As shown in Fig.~\ref{fig.mvddaq} 
a Motorola LynxOS board gathers data through 
the VMEbus from NIKHEF 2TP modules \cite{2tp}, each module consisting of 2 
transputers and a shared triple port memory and providing up to 8 transputer 
links. Data are collected by the CPU, merged event by event and sent through 
Fast Ethernet to the GTT environment.

On GFLT accept digitized pulse height and drift time data from the 4608 CTD 
sense wires are read out from custom FADC cards by 16 sector read-out 
transputers. 
To enable the data from the CTD to be used at the GTT 16 additional 
data splitting TPs were added to the CTD DAQ. These allow data from the FADC 
system to be parasitically read out from the network whilst causing minimal
disruption to the data flow within the CTD network. 

The STT, installed in 2001 and recommissioned in 2003, replace 
the Transition Radiation Detector (TRD) and has been designed to reuse the 
existing TP based read-out system which is similar in design to that of 
the CTD, using the same custom FADC and TP electronics. 
The STT interface to the GTT is implemented using the system developed 
for the CTD, although only 8 input TP links (one 2TP module instead of two) 
are used.
The STT frontend electronics digitise hits above threshold  
and only drift time information is available to the GTT.
The expected STT data volume per GFLT accept is $\leq$4kB with a read-out
latency similar to that of the CTD. At the time of writing the STT
interface to the GTT is fully operational although data was not transferred
during luminosity running to the GTT.

\begin{table*}[htb]
\begin{tabular}{@{}rll}
\hline
Number & Item & Purpose \\
\hline
1 & DELL PowerEdge 6450 Quad 700MHz 1GB	& NFS File Server, EVB Interface and Run Control\\
1 & DELL PowerEdge 4400 Dual 1GHz 256MB	& GTT Server+Credit/GSLT Decision\\
12& DELL PowerEdge 4400 Dual 1GHz 256MB	& GTT Algorithm Processing\\
5 & Motorola MVME2400 450MHz 64MB	& MVD Read-out, CTD and STT Interfaces\\
1 & Motorola MVME2700 367MHz 64MB	& GTT to GSLT Trigger Result Interface\\
1 & Motorola MVME2700 367MHz 64MB	& LynxOS Boot Server and Development Node\\
4 & NIKHEF-2TP VME-Transputer modules   & Transputer Protocol Conversion\\
1 & Motorola MVME2700 367MHz 64MB	& {\tt HELIX} Programming CPU\\
2 & Intel Express 480T Fast/Giga-16Port Cu Switch & Network Connections\\
\hline
\end{tabular}\\
\caption{MVD Data Acquisition and Global Track Trigger 
Computing and Network Resources.}
\label{tab.hw}
\end{table*}

\section{The GTT Environment Process}

The GTT as any of the components participating to the ZEUS second level 
trigger has to process events at the average rate $\le 500$ Hz 
with a mean latency, including all the data transfers, of less than 10 ms 
and possibly small tails due to busy events or performance fluctuations.

Studies of the network throughput using 
commercial PCs and LynxOS PPC boards, indicated, 
after proper tuning of TCP socket options, 
Fast and Gigabit Ethernet to be a reliable low and stable latency connection. 
Feasibility tests using a single data source feeding 4 PCs running dummy 
GTT algorithms on a round-robin basis were made with the data being 
immediately forwarded to a dummy GSLT trigger sink. 
The results in terms of rate and latency indicated the use of TCP/IP 
and Fast Ethernet as acceptable.
Additionally a port of the CTD-SLT TP code to a single CPU showed 
that processing speed was sufficient on PCs to satisfy the 
requirements\cite{gtt}. 

In the current system all MVD DAQ and GTT components are connected using 
the TCP/IP protocol via point-to-point Fast and Gigabit Ethernet links to 
network switches. 

The {\it GTT environment process} is a multi-threaded program with 
one thread per input data source ($3\times$MVD, CTD, STT), one thread 
per trigger algorithm and a time limit thread. 

Typically one environment runs on each of the 12 dual CPU farm PCs.
The location of the next environment to receive event data is
stored in a synchronized ordered list at the data interfaces. When 
available the environment sends its credit to the data source processes. 
The decision to use credits to control availability rather than
round-robin distribution was determined from simulation studies.   

For development and performance tests a {\it playback} capability has been 
provided: upon a First Level Trigger the component front-end electronics 
is read out and Monte Carlo or previously
saved events stored in memory are injected into the GTT trigger chain 
at the component VME interfaces, and sent through the system exactly 
as for regular data.

Currently a {\it barrel} algorithm using CTD+MVD data is implemented. 
A {\it forward} algorithm, using STT+MVD is in preparation.

\section{GSLT and EVB Interfaces}

As the GSLT is based on a transputer network a similar solution 
as for the CTD and the STT was commissioned. 
The GTT result, is sent to a Motorola PowerPC VME CPU which transfers 
the result, via a 2TP module and one of its TP serial links, to the GSLT. 
As the order of the GFLT number of the trigger results arriving at the 
GSLT is strictly sequential the interface is required to order the results 
from the different GTTs before sending.

No special hardware is required to receive the GSLT trigger decision
as this is transferred via TCP. 
An interface process forwards it to the MVD data sources and GTT environment
which processed the event. On accept the data sources send MVD strip
data and the environment sends MVD cluster and
algorithm calculation details to the EVB interface.

The EVB interface waits for MVD and GTT data associated with GSLT
accepted events which are merged, formatted into the final ZEBRA 
and ADAMO banks and sent to the EVB via TCP/IP. A complete data 
quality monitor is performed on this system.

\begin{figure*}[htb]
{\centerline{\psfig{figure=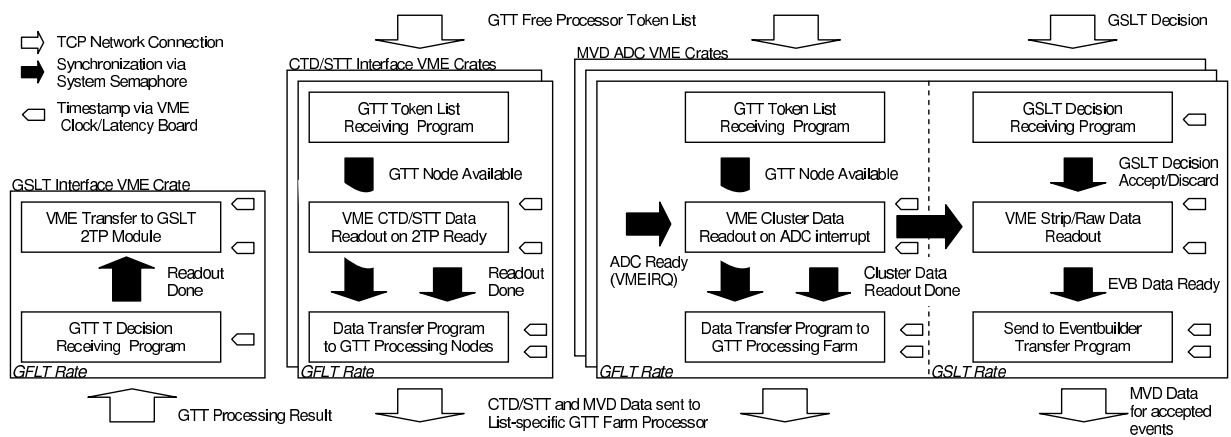,width=16.5cm,clip=}}}
\caption{Block Diagram of the DAQ VME software implementation.}
\label{fig.vmesoft}
\end{figure*}

\section{The VMEbus Access Implementation \label{sec.uvmelib}}

A software package for VMEbus access on Motorola PPC boards 
running LynxOS has been developed. This package
exploits the features of the hardware and the 
operating system to provide flexible VME memory mapping, DMA transfer, 
VME interrupt and process synchronization control in a 
multiuser environment. 
The package consists of a library named {\tt UVMElib}~\cite{uvmelib} 
layered on an enhanced driver for the Tundra Universe II chip 
with respect to the default version distributed by LynxOS. 
The user performs all VME 
and related operations by using the library without any direct 
connection to the driver.  
Within the library, the basic data structure type handled by the user 
to describe Shared Memory Segments opened both on the internal PCI DRAM and/or
on the VMEbus has six fields: an {\it id}, an ASCII {\it name}, a {\it size} 
(in bytes) an addressing {\it mode} and a {\it virtual} and {\it physical} 
address, corresponding to the address 
the application has to use respectively for normal read-write cycles and DMA 
operations. For simplicity both internal contiguous DRAM segments and
VME ones are allocated in the same way.
Standard API are provided to support process 
synchronization by waiting on or setting system semaphores. 
It is possible to connect VME interrupt handling to some UVME semaphores, 
making synchronization to hardware 
interrupts, DMA cycles or software signals equivalent. 
It is worth noting that segments are uniquely identified by 
id or name. This allows many processes to connect to already existing 
mapped regions (up to eight for the VME space) without overloading the system. 

\section{The VMEbus Read-out Software}

The MVD read-out software, due to the strict 
requirements imposed by the participation to the second level trigger, 
has required careful design: the DMA transfer and the VME mapping 
capabilities of the Universe II bridge together with LynxOS specific 
mapping of contiguous shared memories, use of system semaphores and flexible 
priority scheduling have allowed a modular design of the complex DAQ 
environment. 
A diagram of the main processes running on the VME computers when 
taking data is shown in Fig.~\ref{fig.vmesoft} 

On the ADC systems two {\it software pipelines} running 
at FLT and SLT rates exist. At FLT rate the read-out 
program running at lower priority, is woken up on VME interrupt 
as soon as data are ready on the ADC boards. After data has been 
transferred via DMA 
to internal memory, the network tasks, synchonized by a semaphore 
and running at a higher priority, will send the data to the 
GTT farm. The machine destination list is independently updated by 
a network receiving task driven by the GTT process sending its credit. 
A similar software pipeline is available also on the CTD and STT 
interfaces although currently since the 2TP modules cannot generate a 
VMEbus interrupt, the PowerPC polls for data every 500$\mu$s. 

After the GTT processing the trigger result is sent to the GSLT 
and the list of the idle processors is updated. 

All data transfers are performed using standard TCP protocol. 
To cope with the different platforms involved (PowerPC for the 
VME read-out and standard PC for the DAQ and GTT computing nodes) the 
communication and synchronization is done via short XDR encoded 
messages while detector data is sent with no additional overhead. 

To precisely monitor read-out latencies and network transfer 
times an all purpose 6U VME board~\cite{vmeallp}, 
providing among other features 
a {\it latency/clock} register, has been developed and installed in all 
VME crates and connected to a common 16$\mu s$ clock bus. 

Absolute timestamps and latency measurements are available 
for every event and stored in the data at several points 
of the data acquisition phase allowing a complete understanding of 
the system performance. 

\begin{figure}
{\centerline{\psfig{figure=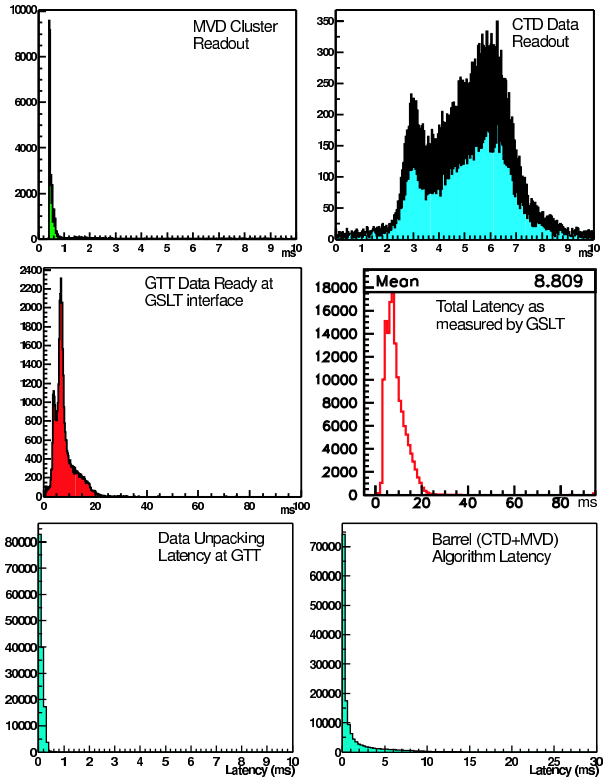,width=8cm,clip=}}}
\caption{\label{fig.lats} Latency measurements from a typical luminosity 
run with high GFLT rate (Nr. 44569) taken in 2003. From top to bottom and 
from left to right the MVD cluster read-out latency (time between ADC FLT 
interrupt and cluster data read-out done), the CTD data read-out completed, 
GTT processing completed the total latency as measured at the GSLT. 
In the lowest row the bare contribution due to data unpacking and 
the processing of trigger algorithm. }
\end{figure}

In Fig.~\ref{fig.lats} the latency distributions for the cluster read-out, 
the CTD data read-out, the algorithm processing and the total GTT system 
latency are shown for a typical luminosity run from 2003. 
A mean value smaller than 10 ms with steep tail, compatible with the 
initial requirements is obtained. 
The data unpacking and algorithm processing latencies for the same run are 
also shown. 
The distribution is monotonically falling, since the algorithm runs as a 
single process, and  has a mean of 1.4~ms with a tail extending to around 
15~ms for busy events. 
A significant contribution to the total latency results from the 
latency of the CTD transputer network in providing data to the CTD interface.
This time is 
not completely wasted as it is used to transfer and unpack the MVD data, 
usually available much earlier. 
For the same run, the average number of credits, i.e. the number of GTT idle 
processors, was 9 with a lower tail of less then 5 occurring for less that 
1\% of the events. This indicates the current farm with 12 GTT nodes as 
more than adequate for the present GFLT rate and total latencies. 

\begin{figure}[tb]
\centerline{\psfig{figure=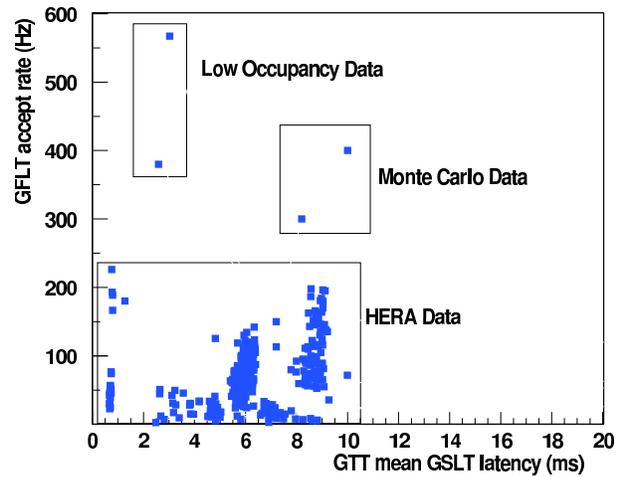,width=8.cm}}
\caption{Mean MVD/GTT Latency versus GFLT rate.}
\label{fig.latvsrate}
\end{figure}

In Fig.~\ref{fig.latvsrate} the GTT total latency versus the GFLT rate 
is shown. The dependence of the mean overall latency on the output 
rate of the GFLT, typically between 50 and 200Hz during this data 
taking period, shows a strong dependence on the background conditions, 
but always lies within 10ms. 

Typical data sizes were 5 kB, 15 kB and 45 kB for CTD, MVD cluster and 
MVD strip total data.

Studies using the playback system, suggest that operating the GTT 
with GFLT output rates of up to 500Hz is achievable.

\section{Computing Environment and Software}

As already described, the DAQ computing environment consists of 
VME boards computers running LynxOS and Intel PCs running Linux 
interconnected via point-to-point Fast or Gigabit Ethernet via network
switches. 
The VME systems are diskless and are booted over the network
from a single, dedicated, VME system acting as a boot server. All
Linux systems are booted from local disk. In order to simplify code
usage a single DAQ file system, containing executable directories
etc., is mounted via NFS by all participating hosts.

Standard C programming has been used throughout the DAQ and GTT systems.
A number of ROOT based C++ GUIs have been developed to 
control the DAQ system in standalone mode, view online or archived 
histograms,  control manually the detector slow control system, etc.

All non read-out or trigger data messages are transmitted through a {\it hub} process 
and not directly between processes. The hub is a multi-threaded processes
which accepts connections at a known address and enforces a simple
protocol, using XDR, which allow connecting
processes to define themselves with known 'public' names or remain 
anonymous and 
set up or cancel forwarding requests. Forwarding
is based on public name, XDR union tag and a 128 bit MD5 hash
field (matches can be made less specific with * wildcards with the public
name to select groups or all names, and 0 when used with tag and hash
fields which enables all matches). 

The hub has a number of other useful
features including retension of the most recent message based on
name-tag-hash. The principle reason for implementing the hub was to
reduce programming complexity in the MVD environment where many
monitoring and logging tasks are running. 
There are, of course, many
tasks running in the system which do not require network
communication.

\subsection{Run and Process Control}

The MVD run control can run in either standalone mode or as part 
of the ZEUS run control system.
In both cases process control, stopping and starting tasks, is facilitated by
daemon processes started at boot time. These advertise themselves to
the run control system and identify, by name, what processes they are 
capable of running, usually this is host specific. 
Run configuration requests from ZEUS or the standalone run control
specify a {\it Runtype} definition file naming all the processes to be 
started of
stopped, the parameters they require, their required exit status etc.
The C preprocessor (cpp) is then invoked to expand all of the
definitions contained in the file (resolving required processes, their
supporting daemon, startup parameters, etc.). The file produced
corresponds to a sequential list of process control commands required
to perform all the transitions specified. Provided no error is
encountered the run control system can then sequence all the steps
required for each transition request received. Transition requests
fail if any process fails to reach and remains in the required state.

\section {The GTT Algorithm}

The current implementation of the GTT algorithm, described in detail 
below relies heavily upon 
the existing CTD-SLT program\cite{ctd} and extends into the MVD 
barrel region. Unlike the CTD-SLT, it does not have access to the CTD 
\zxt\ information and uses instead the data from the CTD stereo 
superlayers for $z$ reconstruction. 
The algorithm consists of four stages: 
\begin{itemize}
\item CTD segment finding, 
\item \rphi track finding (CTD tracks, MVD \rphi hit matching),
\item $z$-track finding (CTD stereo segment matching, MVD $z$ hit matching), 
\item primary vertex identification. 
\end{itemize}
A second pass through one or all stages is performed depending on 
the found tracks and primary vertex, thus improving secondary vertex finding. 

The segment finder operates on both axial and stereo CTD cell data. 
To reduce the processing time, left-right ambiguities, introduced by the 
drift time information and providing real and ghost segment candidates,
are removed by taking the segment pointing more closely to the beam line.
This has a high efficiency for identifying 
high $p_T$ tracks, but leads to a charge asymmetry for lower $p_T$ tracks
due to the $\phi$ asymmetry of the CTD geometry.   
In order to assign hits in a cell to a segment, starting from seed 
pair of hits, the algorithm 
looks for linear segments consisting of three or more hits, using a linear 
extrapolation to identify further hits on adjacent wires until either the 
cell boundary is reached or there are no more hits consistent with the segment.
The segment finder stops looking for
new segments in each cell as soon as at most four segments have been 
found. 
This limit is implemented since the time consumption
is very sensitive to the detector occupancy, which is very high in non $ep$ 
background events. 

\begin{figure}[htp]
\centerline{\psfig{file=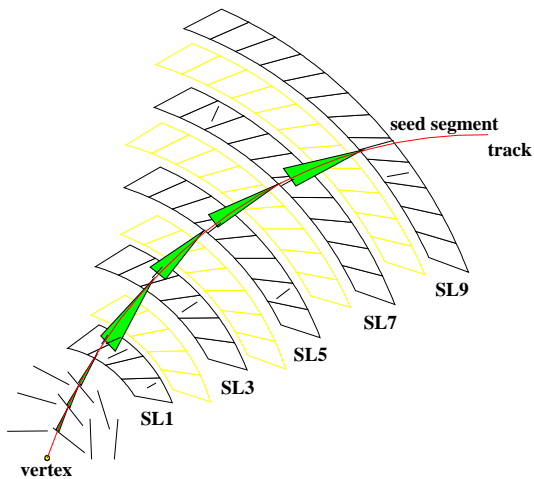,width=7cm}}
\caption{The axial track finding algorithm.} 
\label{axfind}
\vspace{0.5cm}
\end{figure}
The \rphi (axial) track finding, illustrated in Fig. \ref{axfind},
searches for tracks starting with a seed segment in 
the outer superlayer where the occupancy is lowest.
Using this seed segment, the expected azimuthal position of the hit in the 
next innermost axial superlayer is calculated and segments consistent with 
this are matched to the track.
The segment last matched is then used as a fresh seed and the matching proceeds
again into the next inner axial superlayer until at least one segment 
is found in superlayer 1.
Once the segment matching is complete, the track parameters 
are calculated assuming a circle in \rphi using a fast circle fit constrained 
to the beam line to aid subsequent hit matching in the MVD.
Since the MVD hits from both the \rphi and $z$ sensors within an MVD
half module are connected together, all hits must be considered as 
potential \rphi hit candidates. 
Starting from the outermost MVD layer, the $\phi$ position for 
the track at the layer radius is calculated and ladders within a window of 
$\phi$ of this position are considered.
All unmatched hits on these ladders are then considered, calculating the track 
$\phi$ at the radius of each hit. The closest hit on each possible ladder consistent 
with the track is then matched and the track recalculated with the MVD hits
having a larger weight in the fit,  allowing at most
two hits per layer (one per ladder). 
The algorithm then proceeds in the next innermost MVD layer until
all layers contain hits, or there are no further unmatched hits.
When recalculating the track parameters for tracks with at least two 
MVD hits, the 
constraint that the track must come from the beam line is removed, 
in anticipation of secondary vertex finding and impact parameter calculation.

The stereo wire derived $z$ position
$z$ position is only available when the \rphi position of the hit 
on the track has been calculated. 
Since each stereo hit may be assigned to any track passing through the 
large $\phi$ range (4 cells) spanned by the wire, the \rphi and $z$ 
positions of each hit must be calculated for each
possible track candidate within its \rphi range. 
For the innermost stereo layers, this range is up to $36^\circ$, 
which presents a significant problem since the track occupancy nearer 
the interaction region is high and the degree of matching 
ambiguity which must be resolved is large.

The intersection of the track with the hit must be calculated considering 
the drift 
displacement of the hit with respect to the wire. This is done using an 
iterative algorithm \cite{vctrak}
and provides the $\phi$ position of the hit  matched to the track, 
from which the wire $\phi$ position is obtained. 
The fraction of the length along the wire 
is then trivially extracted to provide the $z$ position of the hit. 

Solving the track intersections in \rphi with the stereo wires 
represents one of the most costly steps in terms of the processing 
latency. In order to keep the processing time within acceptable limits, 
as with the axial hit finding, segments are found in the stereo layers 
using the same algorithm as for the axial
layers, with only the end points of the segments used for 
calculating the intersections in $z$. This introduces some additional 
segment finding latency but significantly reduces the time consumption 
in the stereo matching.
\begin{figure}[ht]
\centerline{\psfig{file=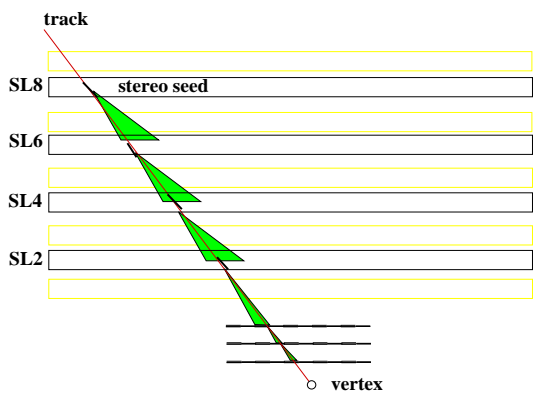,width=7cm}}
\caption{The stereo matching algorithm.}
\label{stereomatching}
\end{figure}
Since the ghost hits lie on the other side of the wire, matching to 
the track yields a similar hit $\phi$ position, but different wire 
$\phi$ position 
and thus a different $z$ position. The ambiguity is resolved by using 
the same beam line requirement as for the axial segments.
Once the segments have been found, 
the stereo matching algorithm proceeds 
starting in the outer stereo superlayer for this track where the 
spatial separation of tracks is highest. All possible segments are considered 
as seed segments. The intersection 
of the segment end points with the track is calculated and fitted  
linearly in $z$ and $s$, the transverse path length along the track, 
to provide a $z$-track.
The $z$-track parameters 
are calculated and the track extrapolated into the next stereo layer. 
The $z$ positions of each segment in the next inner layer are then calculated 
by matching to the track as above.
This continues in each successive stereo layer, with the fit being 
recalculated at each stage, until a track with segments in each stereo layer 
is found, or no matching segment is found.
This is illustrated in Fig.~\ref{stereomatching}.
To improve the efficiency and resolution of track segment 
matching for events where more than a single segment candidate per cell exists, 
the algorithm makes additional passes to find the best candidate. 

Once CTD stereo segments have been matched to an axial track, 
matching to MVD \z-hits is performed in a similar way as in 
the \rphi case. 
Since MVD \rphi positions are already known, the algorithm looks 
for unmatched hits
only in the corresponding \z-sensors of the modules with \rphi hits. 
Starting in the outermost MVD layer, if there are unmatched hits in the 
expected \z-sensor, the track intersection with the sensor is calculated.
All unmatched hits are then compared with the predicted \z-position, and
the closest hit consistent with the track is matched and the track recalculated 
using a higher weight for the MVD hit in the fit.
The calculation of the track intersection and 
hit matching are then successively performed
in the inner layers  
until either hits are found 
in all possible \z-sensors, or there are no hits remaining.
The track-vertex and the weight from the fit are stored for 
use in the primary vertex fit. 

The primary vertex algorithm is intended to make a fast estimation 
of the presence of a possible vertex and to ascertain it's likely position in 
$z$ rather than a complete detailed reconstruction.
The algorithm itself uses a binning algorithm with overlapping 13~cm bins.
The algorithm loops over all tracks, binning the track-vertex intersections 
from the $z$-$s$ fit with the square of the track weight to automatically 
take account of the track quality and the 
better spacial resolution of the MVD.
The most probable bin (MPB) -- the bin with the highest number of weights -- 
is found and  
from the tracks in this bin 
an initial vertex position is calculated using
\[
z_{\rm initial} = \frac{\sum_{i\in {\rm MPB}} z_iw_i^2}{\sum_{i\in {\rm MPB}} w_i^2}.
\]
All tracks within $\pm 9$~cm of this initial vertex are then used to 
calculate the event vertex, again using the weighted mean. 
This has shown to be very stable against the presence 
of incorrectly fitted or assigned tracks.

\section{GTT Performance}

The resolution and efficiency for the track and vertex reconstruction have 
been extensively studied 
using a sample of high transverse energy photoproduction Monte Carlo events
\cite{gttalgo}.
The algorithm is able to adequately reconstruct complex event topologies 
with many tracks with a high precision.  
For tracks found using both
CTD and MVD information, the resolution for the track $z$-vertex position  
is found to be $\sim 500\mu$m
for tracks within the acceptance of the MVD barrel.
The efficiency for track finding rises steeply with the 
track transverse momentum, and is around 80\% for 2D tracks found 
only in \rphi and around 50\% for 
full 3D tracks found in both \rphi and $z$. The loss of efficiency 
when requiring
$z$ information is largely due to the strong pattern recognition 
ambiguities present
from the use of the CTD stereo information and is a strong function of the
track multiplicity in the event, falling from a maximum of around 60\% at 
low multiplicities
to as low as 35\% for track multiplicities of 25, whereas the \rphi
finding efficiency is approximately constant with multiplicity, at 80\%.  

\begin{figure}[tth]
{\centerline{\psfig{file=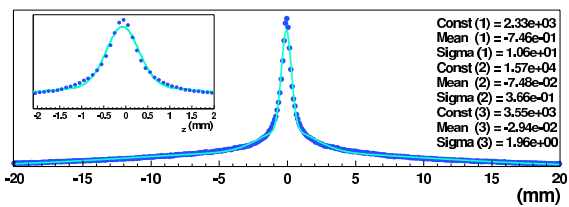,width=7.5cm}}}
\caption{The GTT vertex residual from Monte Carlo data reconstructed 
using CTD and MVD information.} 
\label{mcvertexres}
\end{figure}

The vertex residual with respect to the true position found using the 
dijet photoproduction Monte Carlo sample
is shown in Fig.~\ref{mcvertexres}. The distribution is reasonably well 
described by a sum of three Gaussians, interpolating between  a resolution 
of $\sim 400\mu$m in the central region for event vertices found predominantly 
with tracks including MVD information, and long tails with a resolution of 
around 1cm for vertices found with tracks with 
little or no MVD information.
The efficiency rises rapidly with the track multiplicity and approaches 
100\% for vertices within $\pm 25$cm of the true vertex for events with 
greater than 5 tracks.

During the first commissioning stage of GTT operation with real luminosity 
data, between October 2002 and February 2003, 
the HERA beam gas related background was significantly worse than expected.
To compensate, the CTD had to be operated at only 95\% of the nominal high 
voltage setting leading to a small loss in chamber performance. In addition, 
since beam related background hits in the MVD were seen to bias 
the reconstruction,
the GTT algorithm was running in a mode using 
only CTD information.

\begin{figure}[htb]
\centerline{\psfig{file=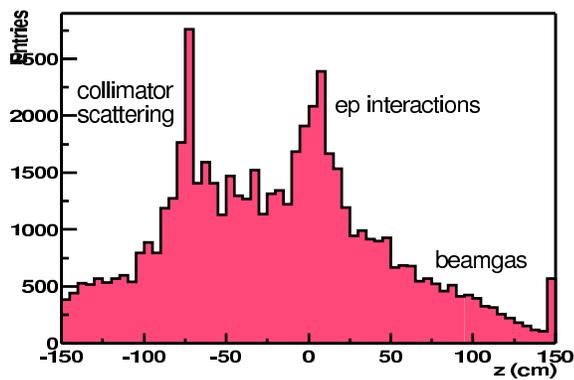,width=7.5cm}}
\caption{The online CTD-only GTT vertex.}
\label{online}
\end{figure}

The performance and stability of the algorithm and GTT system as a whole, was 
well within the expectation during this period. The GTT event vertex available 
online is illustrated in Fig. \ref{online} and clearly shows events from $ep$
interactions in a vertex peak within $\pm 25$~cm, on the large proton-beam gas 
background, together with secondary 
scattering events from the collimator at -80cm.
The efficiency found using GSLT passthrough events
for reconstructing a vertex within $\pm 25$~cm of the offline
vertex is found to be 84\%.

\section{Summary and Outlook}

The MVD DAQ and GTT system have been successfully integrated into the ZEUS
experiment and their performance (latency, stability and efficiency)
are satisfactory. 3.1 million events have been recorded
between Nov/2002 and Feb/2003.

The GTT barrel algorithm performed well during the first commissioning 
luminosity running period of the upgraded
HERA machine, with high stability and latencies
well within those required by the ZEUS DAQ and trigger systems.

The HERA machine is currently undergoing modifications to the interaction 
region to reduce the beam related background. 
The GTT is being improved with the inclusion of a forward algorithm 
and the MVD data will be enabled in the barrel algorithm to
improve tracking and vertex resolution and efficiency. 
These modifications should be available when HERA restarts 
in September 2003.

\section*{Acknowledgment}

The author would like to thank Intel Corp. for the generous 
support in providing the GTT PC-farm and the network hardware. 

The ZEUS MVD and GTT groups are thanked for their
contributions to the DAQ and GTT systems.


\begin{thebibliography}{99}

\bibitem{layout}E. N. Koffeman, \Journal{\NIMA}{\bf 473}{26}{2001}.
\bibitem{helix}M. Feuerstack, \Journal{\NIMA}{\bf 447}{89}{2000}.
\bibitem{adcm}T. Fusayasu, K. Tokushuku, \Journal{\NIMA}{436}{281}{1999}.
\bibitem{uvmelib}A. Polini, ZEUS Note 99-071 (1999), unpublished.
\bibitem{zeus93} ZEUS Collaboration, Status Report, ZEUS 1993, 1997.
\bibitem{ctd} A. Quadt et al., \Journal{\NIMA}{\bf 438}{472}{1999}. 
\bibitem{stt} ZEUS Collaboration, ZEUS Note 98-046 (1998), unpuplished.
\bibitem{ctdint}A. Polini, M. Sutton, S. Topp-Jorgensen ZEUS Note 99-034 (1999), unpublished.
\bibitem{2tp} H. Boterenbrood et al., \Journal{\NIMA}{\bf 332}{263}{1993}. 
\bibitem{gtt}M. Sutton, C. Youngman ZEUS Note 99-074 (1999), unpublished.
\bibitem{vmeallp}A. Polini, M. Pwalam, C. Youngman ZEUS Note 03-009 (2003), unpublished.
\bibitem{vctrak} G. Hartner, ZEUS Note 98-059 (1998), unpublished.
\bibitem{gttalgo}R. Hall-Wilton, M. Sutton, B. West ZEUS Note 99-074 (1999), unpublished.

\end{thebibliography}
\end{document}